\documentclass[aps,prb,twocolumn,showpacs,a4paper]{revtex4}
\usepackage{graphicx}
\usepackage{amsmath}

\newcommand{\bB}{{\bf B}}
\newcommand{\br}{{\bf r}}

\newcommand{\bv}{{\bf v}}
\newcommand{\bF}{{\bf F}} 
\newcommand{\bE}{{\bf E}}
\newcommand{\bd}{{\bf d}}
\newcommand{\e}{\mathrm{e}}
\newcommand{\wc}{{\vec \omega}_c} 
\newcommand{\tom}{\tilde\omega}

\begin{document}

\title{Dynamic response of artificial bipolar molecules}
\author{Egidijus Anisimovas}
\email{egidijus@uia.ua.ac.be}
\author{F. M. Peeters}
\email{peeters@uia.ua.ac.be}
\affiliation{Departement Natuurkunde, Universiteit Antwerpen (UIA),
B-2610 Antwerpen, Belgium} 

\date{27 March 2002}

\begin{abstract}
We calculate the equilibrium properties and the dynamic response of
two vertically coupled circular quantum dots populated by particles
of different electrical charge sign, i.~e.\ electrons and holes. The 
equilibrium density profiles are obtained and used to compute the 
frequencies and oscillator strengths of magnetoplasma excitations. 
We find a strong coupling between the modes derived from the 
center-of-mass modes of the individual dots which leads to an 
anticrossing with a pronounced oscillator strength transfer from 
the ``acoustic'' to the ``optical'' branch. Also, due to breaking 
of the generalized Kohn theorem a number of other than 
center-of-mass modes are excited whose oscillator strengths, 
however, are rather weak.
\end{abstract}


\maketitle


\section{Introduction}
\label{sec_intro}

The physics of quantum dots --- the small man-made structures in a
semiconductor containing anything from a few to thousands or more
electrons --- has already enjoyed a busy and fruitful decade as a 
major field of research in condensed matter 
physics.\cite{jacak98,johnson95,kouwen01} A large part of the 
experimental work on quantum dots focused on the probing of the
electronic states inside the dots by means of far-infrared (FIR)
spectroscopy.\cite{sikorski89,demel90,heitmann93,heitmann95} 
However, a very common feature of many quantum structures is 
their nearly circular shape and a parabolic confining 
potential\cite{shikin91,kumar90} 
which has profound consequences on the optical response. According 
to the generalized Kohn theorem,\cite{brey89} under such conditions
the center-of-mass (CM) motion decouples from the relative motion
of the electrons, and the electric field of the FIR radiation couples 
only to the former. Consequently, the absorption spectra of circular 
parabolic quantum dots consist only of two CM peaks whose positions 
are independent of the electron number and insensitive to the 
electron-electron interaction effects.

Actual experiments performed on arrays of quantum dots containing
a few\cite{sikorski89,meurer92} or up to a few hundred 
electrons\cite{demel90} have confirmed the basic two-peak structure 
in the absorption (transmission) spectra. Small deviations were 
explained by taking into account the nonparabolicity effects,\cite{gudmund91}
lateral Coulomb coupling between the neighboring dots,\cite{chakra92}
and spin-orbit interaction.\cite{jacak98} In a separate line of
development, experimentalists performed a FIR spectroscopy 
study\cite{bollweg96} of the formation of incompressible edge 
stripes\cite{stripes} in quantum dots and antidots with a deliberately
tailored hard-wall confinement, thereby demonstrating that FIR 
spectroscopy is capable of giving a detailed insight into many-body 
systems. As a matter of fact, in this work in order to increase the 
signal strength and create non-parabolic potential profiles, double-layer
dots with three doping layers were used. Therefore, besides the dominant 
``optical'' modes where the electrons  in both layers oscillate in 
phase, also somewhat weaker ``acoustic'' modes were visible.

Recently, the system of two vertically coupled quantum dots was 
also studied theoretically in greater detail.\cite{partoens98} 
This setup is interesting because even in the case when the
confining potentials of the individual dots are parabolic, the
Kohn theorem is broken by the interaction between the non-equivalent 
dots. Therefore, besides the usual CM modes a rather rich spectrum 
of collective modes can be excited, however, most of the oscillator 
strength is still contained in the center-of-mass modes.

In the present paper we study a similar system consisting of two 
vertically coupled circular parabolic quantum dots containing
carriers of opposite charge sign, electrons and holes, respectively. 
Our results are very different from those pertaining to the
previously studied\cite{partoens98} vertically coupled dot system 
and reveal an interesting anticrossing of CM-derived modes 
belonging to separate dots marked by a major oscillator strength 
transfer between them. The kind of system we have in mind can be 
structured in bilayer-bipolar heterostructures containing parallel 
electron and hole layers {\em in equilibrium}. These structures have 
been realized in the crossed gap InAs/GaSb system \cite{inas} as well 
as in biased GaAs/AlGaAs heterostructure \cite{gaas,sivan92} where 
electron and hole layers form on the opposite sides of an AlGaAs 
barrier. Most of the interest in such systems stems from the 
possibility (at least, in principle) of the formation of Bose-Einstein 
condensate of indirect excitons \cite{lozovik99}. While the formation 
of the superfluid state has not been demonstrated so far, a number of 
other interesting effects due to the electron-hole coupling have been 
predicted and/or observed.\cite{sivan92,naveh01} 

Our computational approach is based on the generalization to bilayer
two-component systems of the formalism developed by Zaremba and his
co-workers\cite{zaremba94} and successfully applied to a number of 
electronic systems.\cite{hydro,zyl00} This approach is well suited 
to describe quantum dots with a large number of electrons whose dynamic 
response is dominated by collective excitations.\cite{zyl00} Our paper 
has the following structure. In Section \ref{sec_theory} we discuss
the equilibrium charge-density distribution in artificial bipolar 
molecules. The formalism is given in Sec.\ \ref{sec_dyna}, and the
results regarding the dynamic response of these systems are presented
in Sec.\ \ref{sec_results}. We summarize our results in 
Section \ref{sec_summary}. Two Appendices describe, respectively, the
calculation of Coulomb integrals and a simplified model useful for
obtaining quick estimates of the essential characteristics of the spectrum.

\section{Equilibrium densities}
\label{sec_theory}

We consider two vertically coupled quantum dots, one populated by 
electrons and the other by an equal number of holes. Both electrons
and holes are strictly two-dimensional (2D) and are laterally 
confined by parabolic potentials.

The equilibrium and dynamical properties of a many-electron system
close to the classical regime can be calculated from an approximate 
semiclassical total-energy functional\cite{hydro,zyl00} of the
electron density $n_e$
\begin{eqnarray}
\label{eq_onec}
  E_e[n_e] &=& T[n_e] + \frac{1}{2} \int\! d^2r\, \int\! d^2 r' 
  \frac{n_e(\br) n_e(\br')}{|\br - \br'|} \nonumber\\
  &&+ \int\! d^2 r\, w_e(\br) n_e(\br) + E_{\rm xc}[n_e(\br)].
\end{eqnarray}
Besides the largest contributions of the direct Coulomb interaction
energy and the energy in the external confining potential $w_e$, the
functional (\ref{eq_onec}) includes the quantum-mechanical kinetic
and exchange-correlation energy corrections. Following previous
authors,\cite{hydro,zyl00} we choose to represent the kinetic energy
by its lowest-order gradient (von Weizs\"{a}cker) expansion and 
approximate $E_{\rm xc}$ by the local Dirac exchange-only term. We 
work in the effective atomic units defined by setting 
$\hbar = m_e^{*} = e^2/\epsilon = 1$. Here $m_e^{*}$ is the effective 
{\em electron} mass (the effective hole mass may be different) and 
$\epsilon$ is the dielectric constant of the medium. In these units, 
the kinetic energy and exchange functionals, respectively, are given by
\begin{eqnarray}
\label{eq_explicit}
  T[n_e] &=& \frac{\pi}{2} \int\! d^2 r\, n_e^2(\br) + 
    \frac{\lambda}{8} \int\! d^2 r \frac{|\nabla n_e(\br)|^2}{n_e(\br)},
  \nonumber\\
  \\
  E_x [n_e] &=& -\frac{4}{3} \sqrt{\frac{2}{\pi}} 
    \int\! d^2 r [n_e(\br)]^{3/2},\nonumber
\end{eqnarray}
where $\lambda = 0.25$ is the von Weizs\"{a}cker coefficient.\cite{zyl00}

For a two-component system (electrons and holes) the total energy
consists of the energies of the two subsystems and a coupling term
\[
  E [n_e, n_h] = E_e [n_e] + E_h [n_h] + E_{\rm coup} [n_e, n_h].
\]
The energy functional of the holes $E_h[h_h]$ is identical in 
form to that of electrons as given in Eqs.\ (\ref{eq_onec}) and 
(\ref{eq_explicit}), however, since the holes may have a different 
effective mass the kinetic-energy term is scaled by the inverse 
of the ratio of hole-to-electron effective masses 
$\kappa = m_h^{*}/m_e^{*}$. In our calculations we use the 
characteristic value $\kappa = 3$. (Using the bulk GaAs 
data,\cite{semicond} one obtains $\kappa = 7.9$ and $1.2$ for 
heavy and light holes, respectively.) The inter-layer coupling 
is included at the mean-field level
\begin{equation}
\label{}
  E_{\rm coup}[n_e, n_h] = - \int\! d^2 r_e \int\! d^2 r_h  
  \frac{n_e(\br_e) n_h(\br_h)}{|\br_e - \br_h - \bd|},
\end{equation}
with $\bd$ denoting the vertical separation between the layers.

The equilibrium densities are obtained from the two Euler equations
\begin{subequations}
\label{eq_euler}
\begin{eqnarray}
  \frac{\delta}{\delta n_e} E[n_e,n_h] = \mu_e,\\
  \frac{\delta}{\delta n_h} E[n_e,n_h] = \mu_h.
\end{eqnarray}
\end{subequations}
Following the usual procedure\cite{hydro,zyl00} of evaluating the 
functional derivatives in (\ref{eq_euler}) and expressing them in 
terms of the square root of the particle densities 
$\psi_{e(h)}(\br) = [n_{e(h)}(\br)]^{1/2}$ we arrive at the 
equations determining the equilibrium charge density profiles
\begin{subequations}
\label{sch1}
\begin{eqnarray}
  \left[ -\frac{\lambda}{2} \nabla^2 + u_e(\br) - \mu_e \right]
    \psi_e(\br) &=& 0, \\
  \left[ -\frac{\lambda}{2\kappa} \nabla^2 + 
    u_h({\bf r}) - \mu_h \right] \psi_h(\br) &=& 0. 
\end{eqnarray}
\end{subequations}
with the following expressions for the effective potentials
\begin{subequations}
  \label{sch2}
  \begin{eqnarray}
  u_e({\bf r}) &=& w_e({\bf r}) + \pi \psi_e^2({\bf r}) -
    \sqrt{\frac{8}{\pi}}\psi_e({\bf r}) \nonumber\\ &&+
    \int\! d^2 r' \frac{\psi_e^2({\bf r}')}{|{\bf r} - {\bf r}'|}
    - \int\! d^2 r'\, \frac{\psi_h^2(\br')}{|\br - \br' - \bd|},\\
  u_h({\bf r}) &=& w_h({\bf r}) + \frac{1}{\kappa} \pi \psi_h^2({\bf r}) -
    \sqrt{\frac{8}{\pi}}\psi_h({\bf r}) \nonumber\\ &&+
    \int\! d^2 r' \frac{\psi_h^2({\bf r}')}{|{\bf r} - {\bf r}'|}
    - \int\! d^2 r'\, \frac{\psi_e^2(\br')}{|\br - \br' - \bd|}.
\end{eqnarray}
\end{subequations}
Since the potentials (\ref{sch2}) themselves depend on the solutions
of (\ref{sch1}), the system of equations (\ref{sch1}) and (\ref{sch2}) 
has to be solved self-consistently by convergent iterations. The
angular integrations in Coulomb integrals appearing in Eqs.\ (\ref{sch1}) 
and (\ref{sch2}) can be carried out analytically as described in
Appendix \ref{app_coul}. The remaining radial equations for 
$\psi_{e(h)}$ are solved numerically by discretizing the functions 
and potentials on a grid and using an imaginary-time evolution technique 
described in Ref.\ \onlinecite{zyl00}.

\begin{figure}
\vspace{3mm}
\includegraphics[width=0.45\textwidth]{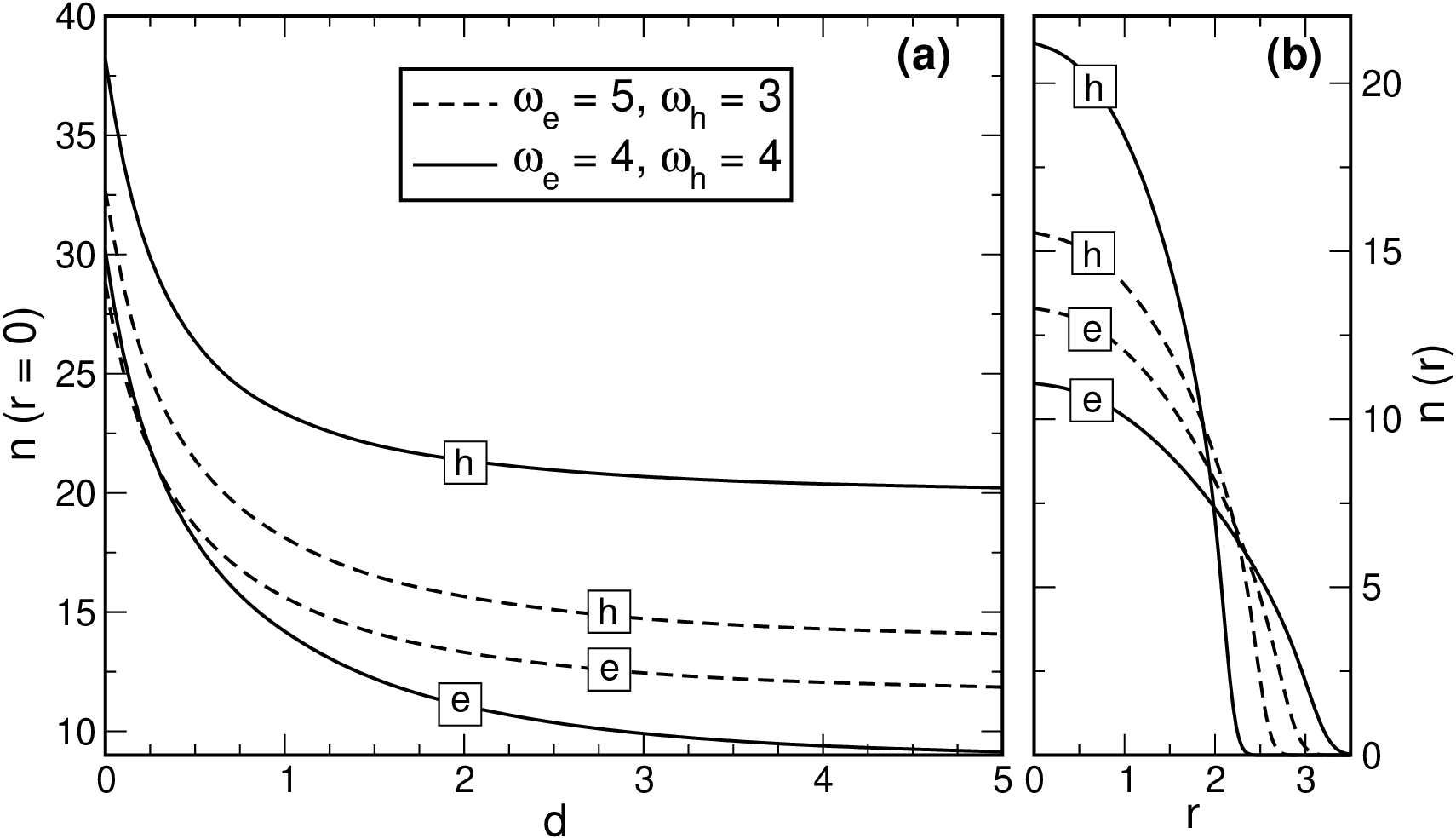}
\caption{\label{fig_density} Equilibrium particle density 
distributions in bipolar quantum-dot molecules. Panel (a) 
depicts the evolution of the particle densities at the center
of the dots versus the inter-dot distance for equal (full line) 
and different (dashed line) confining potentials. Panel (b) 
shows the radial dependences of the electron and hole densities 
for the two situations of (a) at $d=2$.}
\end{figure}

The results obtained using the above formalism are illustrated in 
Fig.\ \ref{fig_density}. In our calculations we use GaAs material
parameters $m_e^{*} = 0.067\,m_e$ and $\epsilon = 13.4$ which 
define the effective Bohr radius (the length unit) 
$a_B^{*} = \hbar^2\epsilon / m_e^{*}e^2 \approx 10$ nm 
and the effective Hartree (the energy unit) 
$E_H^{*} = e^2 / a_B^{*} \approx 10$ meV. Likewise, we express 
frequencies in the
units $E_H^{*}/\hbar \approx 1.5\cdot 10^{13}$ s$^{-1}$, and the
density unit is $a_B^{*\,-2} = 10^{16}$ m$^{-2}$. In panel (b)
of Fig.\ \ref{fig_density}, we show two typical examples of the radial 
distribution of the particle densities in the dots obtained by setting 
the confining potentials to $\omega_e = 5$, $\omega_h = 3$ (dashed line), 
and $\omega_e = \omega_h = 4$ (full line), respectively.
The inter-dot separation is $d = 2$, and each dot contains an equal
number $N_e = N_h = 200$ of particles. We note that near the center 
the density profiles closely follow the usual semi-elliptic 
shape,\cite{shikin91} while at the edges the densities are somewhat 
smoothed if compared to the abrupt square-root behavior predicted 
by the classical treatment.\cite{shikin91,partoens98} This effect is 
due to the included quantum-mechanical corrections which make the 
density to approach zero asymptotically in the classically forbidden 
region.\cite{zaremba94}
Panel (a) displays the particle densities at the central axis 
of the quantum-dot molecule (i.e. the centers of the two dots) as 
a function of the inter-dot distance $d$. Here we use the same values 
of $N_e = N_h = 200$ and two different sets of confining frequencies. 
The full lines are obtained by setting $\omega_e$ and $\omega_h$ to 
the same value $4$; in this case, due to a higher hole effective 
mass the radius of the dot containing holes is smaller and the hole 
densities at the center are considerably larger if compared to the 
electronic dot. The dashed lines illustrate the case of a quantum-dot 
molecule with better balanced radii and densities of its two components. 
Here the confining frequencies are set to $\omega_e = 5$ and 
$\omega_h = 3$. We observe that in all cases the particle densities 
take off rapidly, and hence the dot radii shrink, when the inter-dot 
distance $d$ becomes comparable or smaller than the dot radii which 
are typically in the range $2 \lesssim R \lesssim 3.5$.

\section{Dynamic response -- theory}
\label{sec_dyna}

When the electron-hole system is perturbed away from the equilibrium 
there develops an internal restoring force. The scalar potentials 
of its components $\Phi_{e(h)}$ acting on electron and hole subsystems 
are given by the functional derivatives of the total energy functional 
with respect to the component densities evaluated at the modified 
density values\cite{hydro,zyl00}
\begin{equation}
\label{eq_force}
  \bF_{e(h)}^{\rm int} = - \nabla \Phi_{e(h)}, \,
  \textrm{with } \Phi_{e(h)} = \frac{\delta}{\delta n_{e(h)}} 
  E[n_e, n_h].
\end{equation}
This internal force, along with the external force due to the electric
field of the FIR radiation and the Lorentz force in the presence of a
perpendicular magnetic field, enters the set of four linearized
hydrodynamic equations: the continuity and the force-balance equations 
for each of the two components. However, in order to avoid repetition, 
we will write out explicitly and manipulate only two generic equations
\begin{subequations}
\label{eq_gen}
\begin{eqnarray}
  \frac{\partial}{\partial t} n^1 &+& \nabla \cdot \left[
  n^0 \bv\right] = 0,\\
  \kappa \frac{\partial}{\partial t} \bv &=& - \nabla \Phi + 
  \eta e {\bE} + \eta\bv\times\wc.
\end{eqnarray}
\end{subequations}
The equations describing the hole (electron) layer are obtained from 
(\ref{eq_gen}) by setting the charge-sign factor $\eta = +1$ ($-1$) 
and supplying an extra subscript $h$ ($e$) to the density, velocity 
and potential fields. In addition, for electrons one sets the effective 
mass factor to $\kappa = 1$. In Eq. (\ref{eq_gen}), the Lorentz force 
is expressed in terms of the vector $\wc$ directed perpendicular to
the layers whose absolute value equals the cyclotron-resonance 
frequency of the {\em electrons}.

The quantities $n^0$ and $n^1$ denote, respectively, the equilibrium
density and the linear-order deviation. We work with stationary fields
that depend on time as $\e^{-i\omega t}$ and consider only dipole 
excitations of a given circular polarization. Therefore, in our 
square-root of density notation we express the densities as
\begin{equation}
\label{eq_fields}
  n = \left( \psi + \phi\e^{i\theta} \right)^2,
\end{equation}
here $\theta$ is the angular coordinate. Thus, the two ground-state 
densities equal $n_{e(h)}^0 = \psi_{e(h)}^2$  and the first-order 
fluctuations are given by 
$n_{e(h)}^1 = 2\psi_{e(h)} \phi_{e(h)}\e^{i\theta_{e(h)}}$. Observe 
that the fields $\psi$ and $\phi$ are both circularly symmetric, and 
the correct angular dependence of $n^1$ is explicitly included in the 
factors $\e^{i\theta}$. The external electric field is also taken to be
circularly polarized and derivable from its corresponding scalar potential 
$\bE = - \nabla\Phi_{\rm{ext}}$ with $\Phi_{\rm{ext}} = - Er\e^{i\theta}$
and $E = {\rm const}$. We note that it is {\em not} necessary to consider 
also the opposite polarization $\sim \e^{-i\theta}$ since these results 
can be obtained from the same calculation by simply changing the direction 
of $\wc$, i.e.\ the sign of its vertical projection. Positive (negative) 
values of $\omega_c$ correspond to the direction of the electron (hole) 
cyclotron resonance. Substituting the equilibrium and oscillating densities
from Eq.\ (\ref{eq_fields}) into Eqs.\ (\ref{eq_gen}) we obtain
\begin{subequations}
\label{eq_furt}
  \begin{eqnarray}
  \label{eq_furta}
  &&-i\omega 2 \psi \phi \e^{i\theta} + \rm{div} (\psi^2 \bv) = 0,\\
  \label{eq_furtb}
  &&-i\tom \kappa \bv = - \nabla \Phi +\eta e \bE + \eta \bv \times \wc,
\end{eqnarray}
\end{subequations}
here we also allow for the presence of a small damping force, thus
making a replacement $\omega \to \tom = \omega - i\gamma$ in 
Eq.\ (\ref{eq_furtb}). 

The force-balance equation is readily solved by taking its cross-product
with $\wc$ and using the result to eliminate the cross-product term in
Eq.\ (\ref{eq_furtb}). Straightforward algebra gives
\begin{eqnarray*}
  (\omega_c^2 - \kappa^2\tom^2) \bv &=&
  \left[ i\kappa\tom\nabla\Phi - \eta\nabla\Phi\times\wc\right]
  \nonumber\\
  &&+ \left[ e\bE\times\wc - i\eta \kappa\tom e\bE\right].
\end{eqnarray*}
Substituting this expression into the continuity equation we will need 
only the divergence and the radial component of the velocity field. We 
separate out the angular dependence of $\Phi$ by writing it as 
$\Phi = f(r)\e^{i\theta}$, and carrying out the derivative calculations 
we obtain the wanted quantities
\begin{eqnarray*}
  \lefteqn{ (\omega_c^2 - \kappa^2\tom^2) v_r} \nonumber\\ 
  &=& \e^{i\theta}
  \left( i\tom \kappa f' - \eta \omega_c f \frac{i}{r}  
  - i\eta\tom \kappa e E + i\omega_c e E \right),
\end{eqnarray*}
and
\[
  (\omega_c^2 - \kappa^2\tom^2) {\rm div } \bv = i\tom \kappa \nabla^2\Phi
  = i\tom \kappa \e^{i\theta} \Delta_r f.
\]
Here the operator
\[
  \Delta_r = \frac{d^2}{dr^2} + \frac{1}{r} \frac{d}{dr} - \frac{1}{r^2}
\]
denotes the radial part of the Laplacian.

Putting everything together we end up with the following generic 
equation for the charge density fluctuation $\phi$
\begin{eqnarray}
\label{eq_solved}
  &-&\omega (\omega_c^2 - \kappa^2\tom^2) \phi + 
  \tom \kappa \left( f' \psi' + \frac{1}{2}\Delta_r f \psi \right)
  - \omega_c \eta f \frac{1}{r} \psi'\nonumber\\ &&=
  eE\psi' (\eta\tom \kappa - \omega_c)
\end{eqnarray}
which will generate the two equations for the electron and hole layers. 
We stress that the quantities $f$ (radial parts of the potentials 
$\Phi$) themselves are linear functionals of $\phi$'s. Therefore, 
the right hand side of Eq.\ (\ref{eq_solved}) can be represented as a 
result of an application of a certain linear operator on $\phi$. 

We write the internal potentials $\Phi_{e}$ and $\Phi_{h}$ as
\[
  \Phi_e = \Phi_{ee} + \Phi_{eh}, \textrm{ and }
  \Phi_h = \Phi_{he} + \Phi_{hh},
\]
thus separating the internal potentials created by inter- and intra-layer 
interactions. The expressions of the respective contributions are obtained 
by straightforward functional-differentiation from the definition 
(\ref{eq_force}) and read
\begin{subequations}
\label{yahoo}
\begin{eqnarray}
  \Phi_{ee} &=& 2\pi\psi_e\phi_e - \frac{\lambda}{2} \psi_e^{-2}
    (\psi_e \nabla^2 \phi_e - \phi_e \nabla^2 \psi_e) 
    \nonumber\\ &&- \sqrt{\frac{8}{\pi}} \phi_e
    + 2\int\! d^2 r'\, \frac{\psi_e(\br')\phi_e(\br')}{|\br-\br'|},
    \\ 
  \Phi_{hh} &=& \frac{2\pi}{\kappa} \psi_h\phi_h - 
    \frac{\lambda}{2\kappa} \psi_h^{-2}
    (\psi_h \nabla^2 \phi_h - \phi_h \nabla^2 \psi_h) 
    \nonumber\\ &&-   \sqrt{\frac{8}{\pi}} \phi_h
    + 2\int\! d^2 r'\, \frac{\psi_h(\br')\phi_h(\br')}{|\br-\br'|},
    \\
  \Phi_{eh} &=& 
    - 2\int\! d^2 r'\, \frac{\psi_h(\br')\phi_h(\br')}{|\br-\br' - \bd|},
    \\ 
  \Phi_{he} &=& 
    - 2\int\! d^2 r'\, \frac{\psi_e(\br')\phi_e(\br')}{|\br-\br' - \bd|}.
\end{eqnarray}
\end{subequations}
The Coulomb integrals entering the expressions in Eqs.\ (\ref{yahoo})
resemble those encountered in the calculation of the equilibrium
properties in Eqs. (\ref{sch1}) and (\ref{sch2}). However, in the 
present case we deal with $p$-wave charge distributions of angular
dependence $\sim \e^{i\theta}$ creating $p$-wave electrostatic
potentials. The calculation of these integrals is also discussed 
in Appendix \ref{app_coul}.

For the sake of notational compactness, we introduce linear operators 
$\cal L$ corresponding to the different terms in these expressions, so 
that Eq.\ (\ref{eq_solved}) can be written as
\begin{subequations}
\begin{eqnarray}
  -\omega (\omega_c^2 - \tom^2) \phi_e +
    {\cal L}^{ee} \phi_e + {\cal L}^{eh} \phi_h &=& R^e,\\
  -\omega (\omega_c^2 - \kappa^2\tom^2) \phi_h +
    {\cal L}^{he} \phi_e + {\cal L}^{hh} \phi_h &=& R^h,
\end{eqnarray}
\end{subequations}
where $R^e = -eE\psi' (\tom + \omega_c)$ and 
$R^h = eE\psi'(\tom\kappa - \omega_c)$ stand for expressions on the 
right-hand side of Eq.\ (\ref{eq_solved}).

We choose to expand the fields $\phi_e$ and $\phi_h$ in the set
of Darwin-Fock functions of the angular momentum $M = 1$. Thus,
\begin{subequations}
\label{eq_basis}
\begin{eqnarray}
  \phi_e &=& \frac{\e^{i\theta}}{r_0} \sum_{n=0}^{\infty} a_n g_n (r/r_0),\\
  \phi_h &=& \frac{\e^{i\theta}}{r_0} \sum_{n=0}^{\infty} b_n g_n (r/r_0)
\end{eqnarray}
\end{subequations}
with $a_n$ and $b_n$ being the expansion coefficients and the radial
functions
\[
  g_n (r) = \sqrt{\frac{2}{n+1}}\,\e^{-r^2/2}\,rL_n^1(r^2)
\]
are expressed in terms of the associated Laguerre polynomials $L_n^1(x)$. 
The expansions (\ref{eq_basis}) can be optimized by tuning the scaling 
radius $r_0$, and typically some $20$ terms are needed in 
Eqs.\ (\ref{eq_basis}) to obtain convergent results.

The remaining task is the numerical calculation of the matrix elements 
of the operators ${\cal L}$ in the basis (\ref{eq_basis}) leading to 
the coupled set of linear equations
\begin{subequations}
\begin{eqnarray}
  - \omega(\omega_c^2 - \tom^2) a_n 
    + \sum_{n'} ({\cal L}_{nn'}^{ee} a_{n'} 
    + {\cal L}_{nn'}^{eh} b_{n'}) = R_n^e,\quad &&\\ 
  - \omega(\omega_c^2 - \kappa^2 \tom^2) b_n 
    + \sum_{n'} ({\cal L}_{nn'}^{he} a_{n'} 
    + {\cal L}_{nn'}^{hh} b_{n'}) = R_n^h,\quad &&
\end{eqnarray}
\end{subequations}
which we solve numerically by LU decomposition\cite{nrc} 
and obtain the sets of 
expansion coefficients $a_n$ and $b_n$. This enables us to reconstruct 
the fluctuating charge density profiles $\phi_{e(h)}$ in the two layers 
and evaluate the energy dissipation due to the Joule heating
\begin{equation}
  P_{e(h)} (\omega) = \mp2\pi eE \omega\, {\rm Im} 
  \left[\int_0^{\infty}\!\! dr\, r^2 \psi_{e(h)} \phi_{e(h)} \right].
\end{equation}
The sum $P = P_e + P_h$ determines the absorption rate.

\section{Dynamic response -- numerical results}
\label{sec_results}

\begin{figure}
\vspace{3mm}
\includegraphics[width=0.45\textwidth]{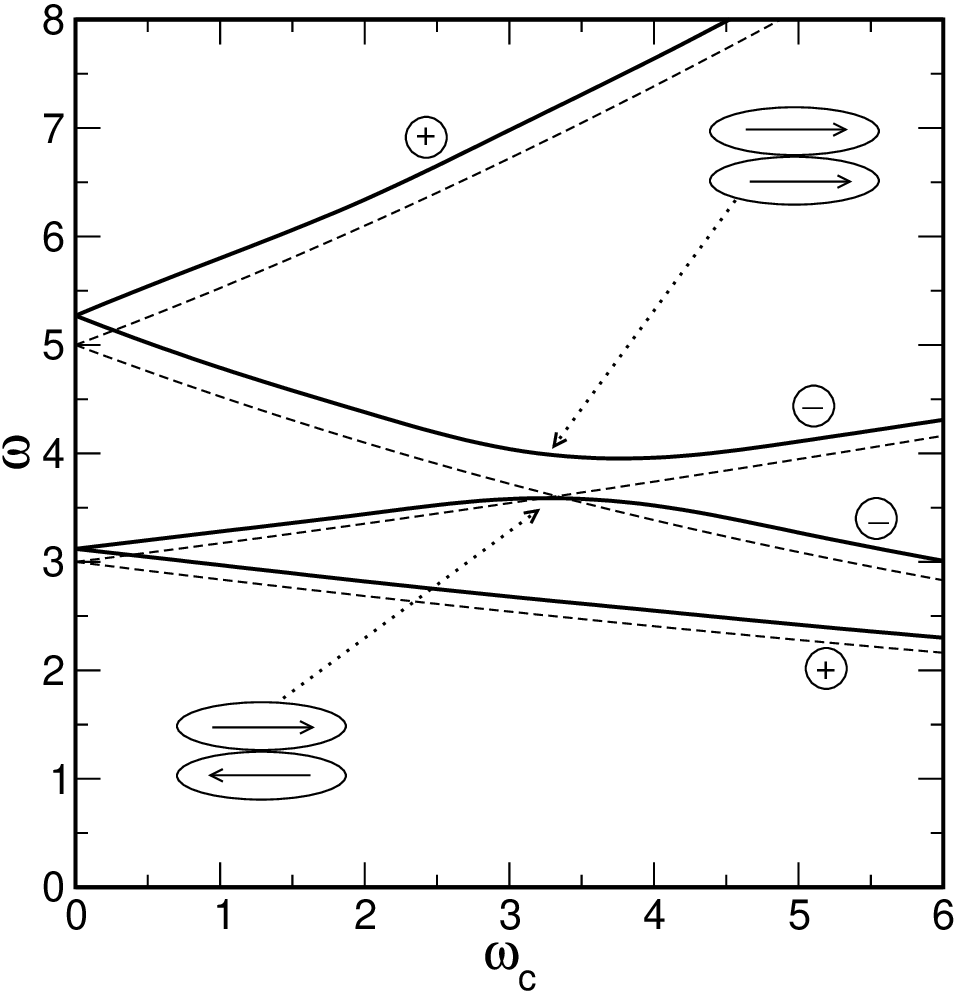}
\caption{\label{fig_intro}The magnetic field dependence of the four 
principal modes in the spectrum of a bipolar quantum-dot molecule. 
Symbols ``$+$'' and ``$-$'' indicate the polarisations. Full lines 
correspond to the vertical separation of $d = 3$, and dashed lines 
depict the decoupled limit $d = \infty$. The insets illustrate the 
relative arrangements of electrical dipoles of the two dots pertinent 
to the respective anticrossing branches.}
\end{figure}

Turning to the description of the absorption spectra of coupled bipolar 
quantum dots, we begin by discussing the four most conspicuous modes 
which evolve from the centre-of-mass (CM) modes of the two individual 
dots henceforth referred to as CM modes. Later, we proceed to describe
the higher resonances and low-frequency edge modes whose oscillator 
strengths are inherently weak thus rendering them more difficult to 
observe experimentally.

\begin{figure}
\vspace{3mm}
\includegraphics[width=0.45\textwidth]{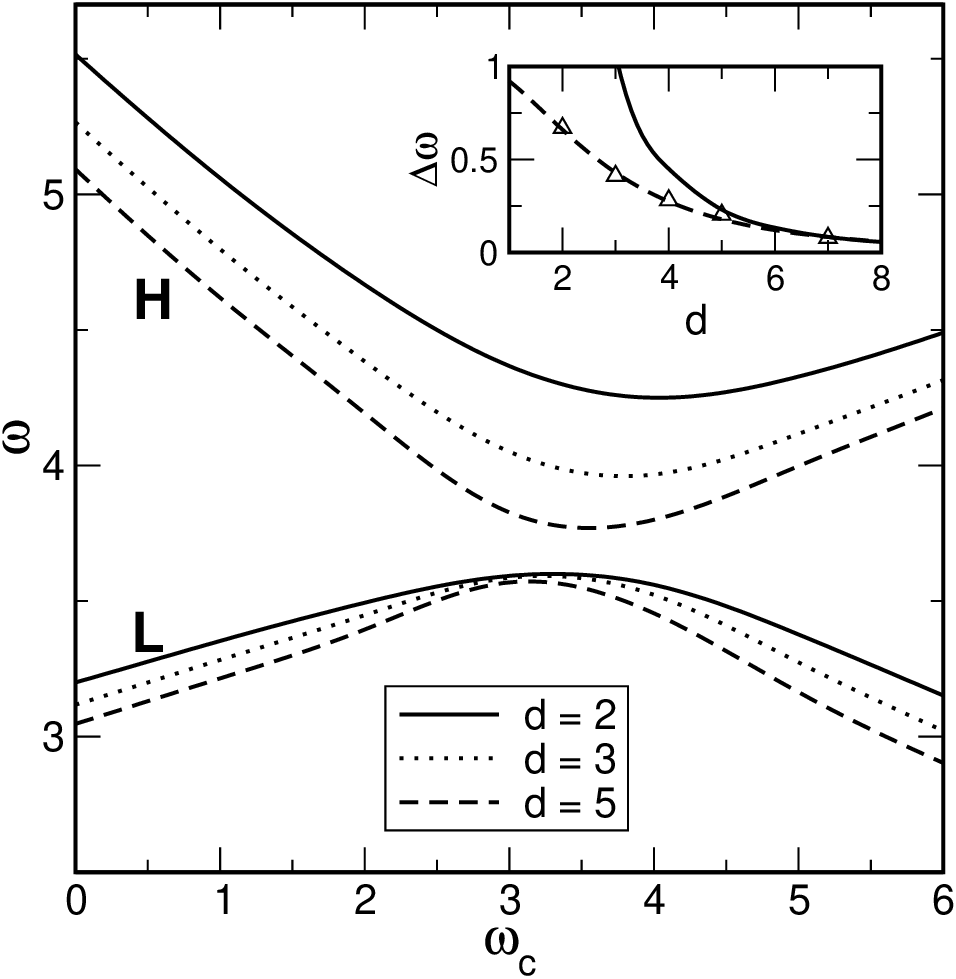}
\caption{\label{fig_cmw}The magnetic-field dispersion of the 
anticrossing CM modes of negative polarisation plotted for three 
different values of the vertical inter-dot separation $d$. The 
symbols ``H'' and ``L'' label the higher and the lower 
anticrossing branches. Note the widening of the anticrossing gap 
and its shift towards higher frequencies with decreasing $d$.
The inset compares the gap-widths obtained from a numerical 
calculation (triangles) and its fit (dashed line) to those obtained 
from the simplified model of Appendix \ref{app_model} (full line).}
\end{figure}

\begin{figure}
\vspace{3mm}
\includegraphics[width=0.45\textwidth]{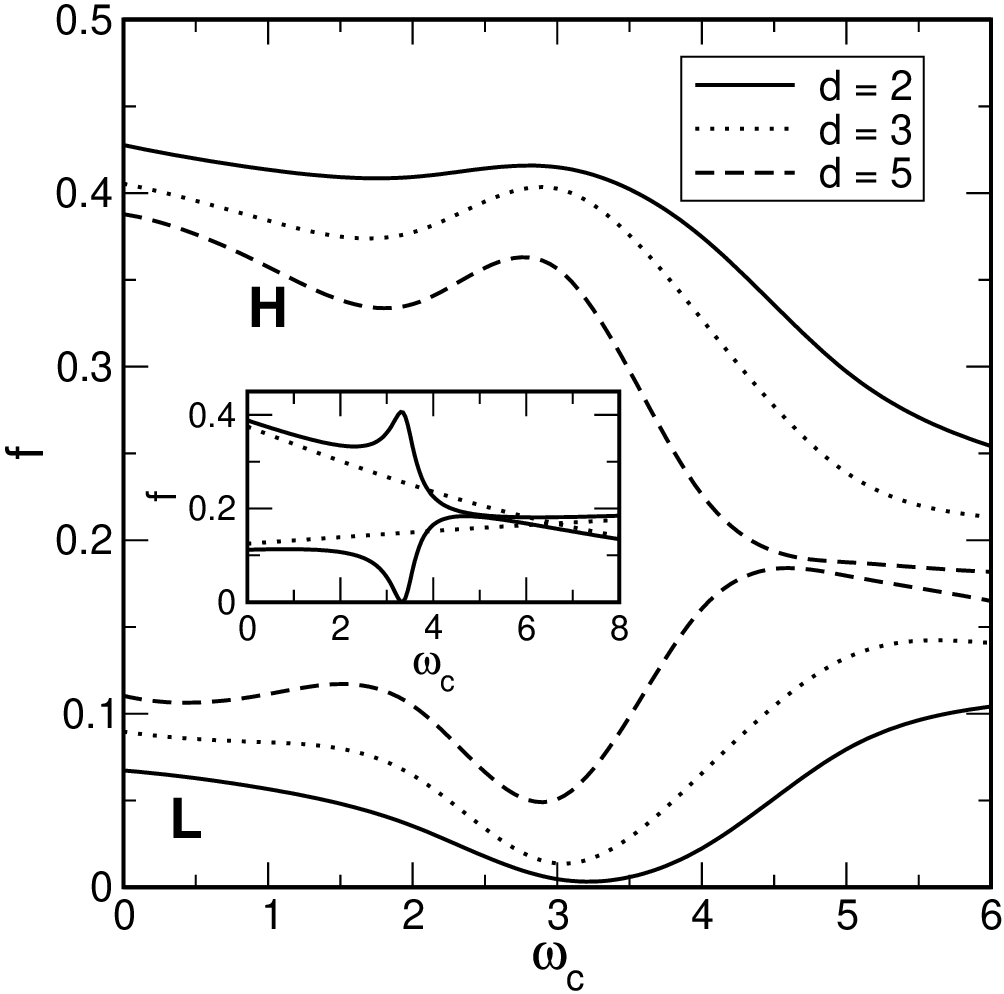}
\caption{\label{fig_cmf}The oscillator strengths of the two anticrossing 
CM modes of the negative polarization. The upper (lower) family of curves 
corresponds to the ``optical'' (``acoustic'') branch. A pronounced 
oscillator strength transfer between the modes in the magnetic-field 
range $2 < \omega_c < 4$ is apparent. As in Fig.\ \ref{fig_cmw}, the 
symbols ``H'' and ``L'' indicate the higher and the lower anticrossing 
branches, respectively. The inset depicts the results obtained from
the simplified model described in Appendix \ref{app_model}. The full
(dotted) lines correspond to coupled (uncoupled) dots.}
\end{figure}

The basic structure of the CM-mode spectrum as a function of magnetic 
field is shown in Fig.\ \ref{fig_intro}. These results are obtained for 
coupled dots containing $N_e = N_h = 200$ particles each with confinement 
frequencies set to $\omega_e = 5$ and $\omega_h = 3$ (as in the second 
example of Sec.\ \ref{sec_theory}). The vertical separation between the 
dots is $d = 3$. The encircled symbols ``$+$'' and ``$-$'' designate 
the directions of the circular polarizations of the respective modes. 
Our convention is to take the direction of the {\em electronic} cyclotron 
resonance as ``positive'' (``$+$'') and vice versa. The narrow dashed 
lines indicate the positions of the CM modes of decoupled dots 
(i.~e.\ $d = \infty$). In agreement with the generalized Kohn theorem, 
at zero magnetic field their frequencies coincide with the confinement 
frequencies $\omega_e$ and $\omega_h$, and split into two branches at 
finite magnetic fields. One notes that due to the Coulomb coupling between 
the dots: (i) all the modes are slightly displaced upwards with respect 
to their positions at $d = \infty$, and (ii) the two middle modes which 
are polarised in the same ``$-$'' direction anticross, while the two 
modes of ``$+$'' polarisation reside in distinct frequency regions and 
thus interact only very weakly. Both points indicate {\em significant 
differences} from the spectra of vertically-coupled electronic quantum 
dots\cite{partoens98} where the inter-dot coupling induces shifts of 
the modes towards {\em lower} frequencies and no such anticrossing is 
observed. In these systems, the interaction only couples pairs of CM
modes that both have positive or negative magnetic-field dispersion
and do not cross in the absence of interaction. Thus we see that the 
charge-sign reversal of particles in one of the dots does indeed 
induce a substantial qualitative difference. In Appendix 
\ref{app_model}, we show that essential features of the CM mode 
spectrum can be captured within a simplified coupled harmonic-oscillator 
model which can be useful in obtaining quick estimates.

We take a closer look at the anticrossing modes in Figs.\ \ref{fig_cmw} 
and \ref{fig_cmf} which show, respectively, the behaviour of the 
frequencies of the two anticrossing branches and their oscillator 
strengths for three different values of $d$. We note  from 
Fig.\ \ref{fig_cmw} that as the inter-dot separation becomes smaller 
and coupling between the dots increases, the anticrossing becomes more 
pronounced while at the same time both branches tend to shift towards 
higher frequencies. The inset of Fig.\ \ref{fig_cmw} shows the
dependence of the size of the anticrossing gap $\Delta\omega$ on 
the inter-dot separation $d$. The triangles denote the actual 
calculated values while the full line shows the result obtained
from a simple harmonic-oscillator model discussed in Appendix
\ref{app_model}. This model predicts that the gap grows proportionally
to $\Delta\omega \sim d^{-3}$, however, this law is valid only at 
relatively large distances $d > 5$. We found that for smaller values
of $d$ the splitting could be reasonably well fitted by the dependence
$\Delta\omega \sim (d^2 + d_0^2)^{-3/2}$ indicated by a dashed line 
in the inset of Fig.\ \ref{fig_cmw}.

The evolution of the respective oscillator strengths in Fig.\ 
\ref{fig_cmf} is rather peculiar and requires a more detailed
explanation. The sum of the oscillator strengths satisfies a
sum rule, are we normalise it so that their total sum equals $1$. 
In the present example the electron 
and hole numbers are set equal while the holes are taken to be 
$\kappa = 3$ times heavier than the electrons. Therefore, since the
oscillator strengths scale as $\sim N/m$, the electronic modes possess 
$3$ times higher oscillator strengths. Thus, at zero magnetic field the 
oscillator strengths of the two anticrossing modes start from the values 
close to $0.125$ for the lower-energy mode which is essentially localised
in the hole subsystem and $0.375$ for the higher one. These numbers are
slightly modified due to the interaction between the modes (the role of 
interaction becomes more important at lower values of $d$) as well as 
due to the presence of other much weaker modes. The rest $50~\%$ of the 
total oscillator strength at $\omega_c = 0$ belongs to the modes of the 
opposite ``$+$'' polarisation not shown here.

\begin{figure}
\vspace{3mm}
\includegraphics[width=0.45\textwidth]{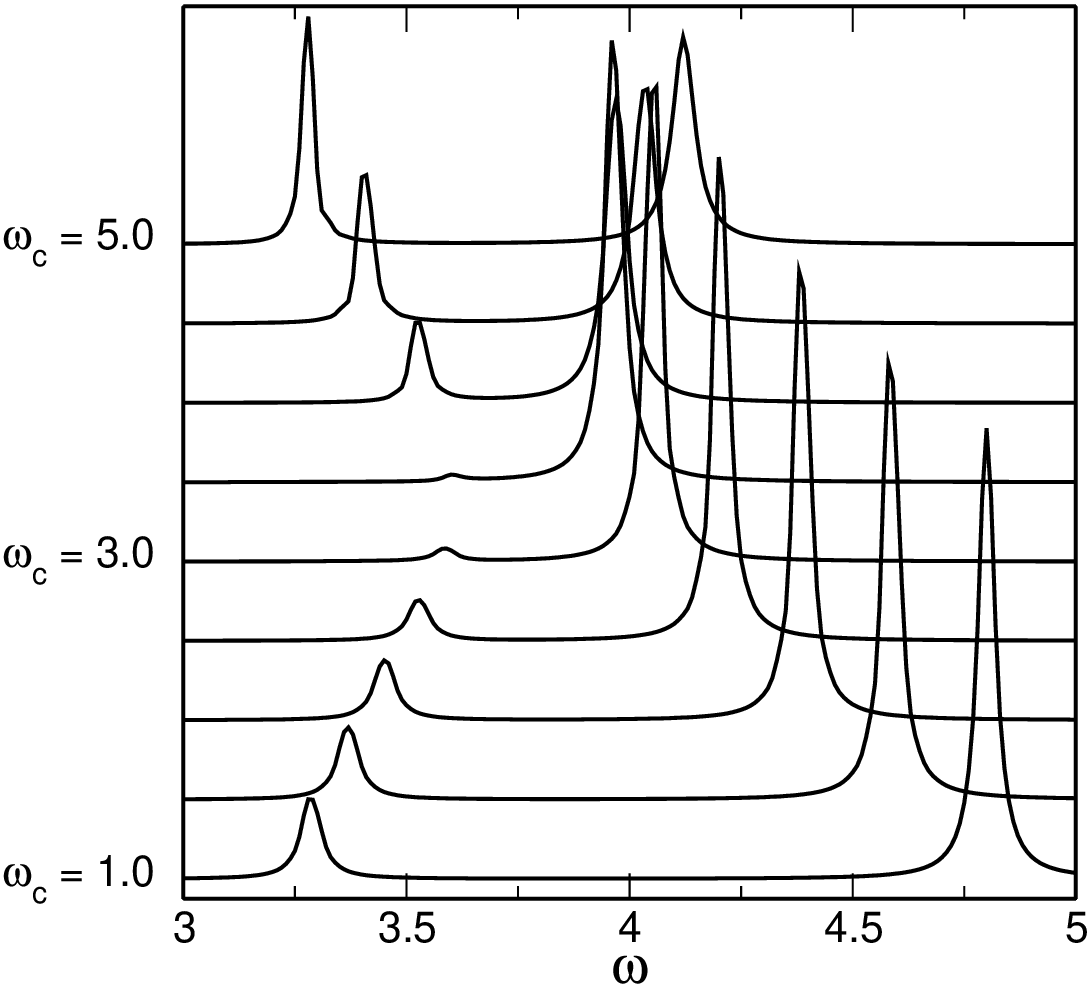}
\caption{FIR absorption (in arbitrary units) in a bipolar quantum-dot 
molecule close to the anticrossing region. We show nine absorption 
curves corresponding to evenly spaced magnetic field values and offset 
vertically by the same amount for clarity. Note the ``disappearance'' 
of the lower branch. A broadening $\gamma = 0.05$ has been used.}
\label{fig_peaks}
\end{figure}

In the range of magnetic-field strengths $2 < \omega_c < 4$ the
two modes interact strongly and anticross. In this region, one observes
a rather pronounced depression in the oscillator strength of the lower
branch reaching nearly zero value. The missing oscillator strength is
transferred to the high-frequency branch. This type of behaviour can 
be understood by realising that at the anticrossing point the charge
density oscillations of the individual dots combine together in either
``optical'' (the two electric dipoles being aligned in parallel) or 
``acoustic'' (antiparallel dipoles) fashion, as illustrated in the
insets of Fig.\ \ref{fig_intro}. Naturally, the parallel alignment of 
two dipoles costs more energy, and therefore this ``optical'' mode 
has a higher oscillation frequency, while at the same time it possesses
a larger net dipole moment, and consequently, a higher oscillator 
strength. As one notes in the inset of Fig.\ \ref{fig_cmf}, the same
type of qualitative behaviour is also observed in the coupled
harmonic-oscillator model of Appendix \ref{app_model}. However, as
far as oscillator strengths are concerned, its quantitative 
predictions are not trustworthy, and thus one can only rely on the 
more-accurate numerical treatment. Our calculations predict the above 
described oscillations of the oscillator strengths to be quite 
strong. In Fig.\ \ref{fig_peaks} we show a set of absorption lines 
simulating those obtainable in FIR spectroscopy 
measurements\cite{jacak98,heitmann93,bollweg96} which have been 
calculated for the case of a vertical separation between 
the dots $d = 3$. We plot nine lines corresponding to nine equally 
spaced values of the magnetic fields between $\omega_c = 1$ and 
$\omega_c = 5$ (the anticrossing region) thus making the ``disappearance'' 
of the low-frequency branch apparent. Since the fluctuations of the 
oscillator strengths of these modes can be rather large (of order of 
$10~\%$ of the total oscillator strength) we expect that the above 
described effect could be readily observed experimentally. It is worth
mentioning that while in the above examples we always dealt with equal 
numbers of electrons and holes in the dots, the ratio of the electron 
and hole numbers can be useful to balance the distribution of oscillator 
strengths between the modes. For example, increasing the number of holes 
will brighten the modes that are mostly due to oscillations in the hole 
subsystem thereby compensating for their diminished oscillator strengths 
because of higher hole effective mass. As we show in 
Appendix \ref{app_model}, the strength of the interaction between the 
two dots and the anticrossing gap scale as the geometric mean of
the electron and hole numbers.

\begin{figure}
\vspace{3mm}
\includegraphics[width=0.45\textwidth]{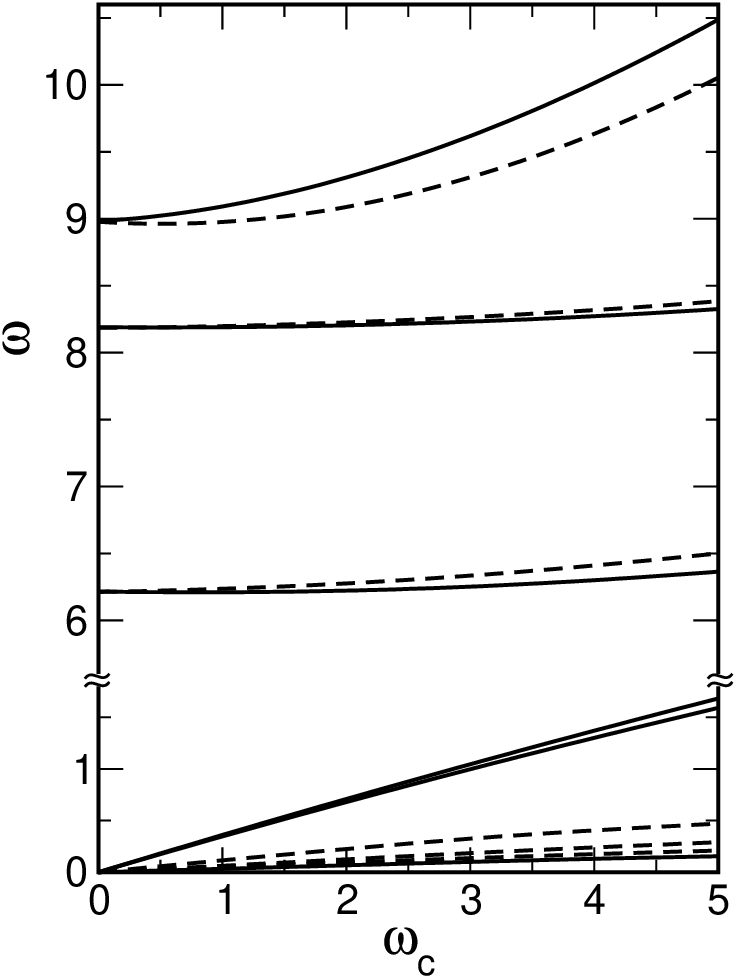}
\caption{The magnetic-field dispersion of the higher-resonances and 
edge modes in a bipolar quantum-dot molecule. Note that the middle 
part of the spectrum is cut out. The full and dashed lines denote 
the modes of the positive and negative polarisation, respectively.}
\label{fig_hilo}
\end{figure}

Besides the above described strong modes the spectra of artificial
molecules feature a number of other rather weak modes that can be 
classified into higher resonances and low-energy edge modes. Due
to a stronger localisation of the charge-density oscillations
associated with these modes, the number of terms in the expansions
(\ref{eq_basis}) has to be increased for an accurate representation.
The oscillator strengths of these modes, under the conditions of our 
calculations, typically barely reach $10^{-4}$ of the total sum of 
the oscillator strengths. 
In Fig.\ \ref{fig_hilo}, we show the magnetic field dispersions 
of the lowest lying high-energy modes and several strongest edge modes. 
These results are obtained for the parameter values $N_e = N_h = 200$,
$\omega_e = 4$, $\omega_h = 3$, and $d = 5$. The full (dashed) lines 
denote branches polarised in the positive (negative) direction. In general, 
the locations and dispersions of these modes resemble analogous 
modes as predicted for coupled quantum dots populated solely by 
electrons.\cite{partoens98} 
The key difference is that in the present case one can classify the modes 
into those dominated by oscillations in either the electron or hole 
subsystems. Thus, the two lower high-energy resonances in 
Fig.\ \ref{fig_hilo} are mostly due to holes while the topmost mode is
mainly electronic. The difference can be most easily spotted in the 
relative arrangement of the positively and negatively polarised branches
at finite magnetic fields. In the electronic mode, the upper branch is
polarised in the positive (i.~e.\ electronic cyclotron resonance) direction,
while in the case of the other two high-energy modes the situation is 
reversed. Moreover, the electronic mode features a considerably stronger 
magnetic-field dispersion. This is due to the fact that the higher 
resonances asymptotically approach the cyclotron-resonance line which 
is $\kappa = 3$ times steeper for electrons than for holes.

The lower part of Fig.\ \ref{fig_hilo} shows six most conspicuous
edge modes of the bipolar quantum-dot molecule. The frequency range
covered by these modes and their dispersion again resemble the case 
of electronic coupled dots.\cite{partoens98} However, due to the 
fact that the polarisations of the edge modes are determined by the 
charge-sign of the carriers, in the present case the edge modes can 
be polarised both in the cyclotron and the anticyclotron directions.
Moreover, the direction of polarisation of an edge mode also betrays 
the component whose contribution is dominant. Thus, the edge modes 
mostly influenced by the electronic subsystem are polarized in the 
``$-$'', i.e.\ {\em anticyclotron} direction of the electrons. These 
modes a depicted by the dashed lines in Fig.\ \ref{fig_hilo}. 
Contrariwise, the full lines in Fig.\ \ref{fig_hilo} denote the 
modes mostly due to the oscillations in the hole subsystem which 
are polarised in the ``$+$'', i.~e.\ electronic cyclotron 
resonance direction.

\section{Summary}
\label{sec_summary}

In conclusion, we made a theoretical investigation of the 
equilibrium density distributions and the far-infrared response 
of bipolar quantum-dot molecules within a hydrodynamic model 
including the effects due to exchange and kinetic-energy in 
the von Weizs\"{a}cker approximation. The most conspicuous 
effect we found is the pronounced anticrossing between two modified 
center-of-mass modes which takes place when the applied magnetic 
field aligns the frequencies of two centre-of-mass modes. The 
additional distinguishing feature of this anticrossing is the 
strongly non-monotonous behaviour of the oscillator strengths of 
the two resulting branches. The oscillator strength transfer from 
the lower ``acoustic'' to the higher ``optical'' branch is certainly
strong enough to be easily observable. On the other hand, the higher 
resonances and edge modes which are also excited in the considered 
setup are quite weak.

\acknowledgments

This work is supported by the Inter-University Attraction Poles
(IUAP-V) program, concerted action (GOA) and the University of 
Antwerp through a VIS-project. E.A. is supported through a
European Community Marie Curie Fellowship.

\appendix

\section{Coulomb integrals}
\label{app_coul}

In this problem we encountered two types of Coulomb integrals
\begin{eqnarray}
\label{eq_jsp}
 J_s (r) &=& \int\!\! d^2 r' \frac{\psi_0^2(r')}
   {|\br - \br' -\bd|},\nonumber\\
  \\
 J_p (\br) &=& 2\int\!\! d^2 r' \frac{\psi_0(r')\psi_1(\br')}
   {|\br - \br' -\bd|}.\nonumber
\end{eqnarray}
Here, $\psi_0(r)$ depends only on the radial coordinate $r$ while
$\psi_1(\br) = f_1(r)\e^{i\theta}$ has a $p$-wave angular dependence. 
$d$ is the vertical separation between the layers which as a special
case may equal zero.

The angular integration in of Eq.\ (\ref{eq_jsp}) can be
carried out analytically in terms of the complete elliptic functions
of the first kind $K(k)$, and the second kind $E(k)$. Introducing
$\rho^2 = (r + r')^2 + d^2$ we obtain
\begin{eqnarray}
  J_s(r) &=& 4\int_0^{\infty}\!\! dr' r' \frac{\psi_0^2(r')}{\rho}
    \,K\!\left(\frac{\sqrt{4rr'}}{\rho}\right), \nonumber\\
  \\
  J_p(\br) &=& 8\e^{i\theta} \int_0^{\infty}\!\! dr' r'
    \frac{\psi_0(r')f_1(r')}{\rho} 
    \,C\!\left(\frac{\sqrt{4rr'}}{\rho}\right),\nonumber
\end{eqnarray}
with
\[
  C(k) = \frac{2}{k^2}[E(k) - K(k)] - K(k).
\]

\section{Coupled harmonic-oscillator model}
\label{app_model}

Guided by a similar simple model introduced in 
Ref.\ \onlinecite{partoens98}, we show that a number of basic features 
of the CM-mode spectrum can be 
derived (at least qualitatively) from a simplified model featuring two 
coupled harmonic oscillators. The oscillators concentrate the total masses 
and charges of the two coupled dots and interact via the potential
\begin{equation}
\label{eq_b1}
  \frac{N_e N_h}{\sqrt{d^2 + (\br_e - \br_h)^2}} \approx
  - \frac{N_e N_h}{d} + \frac{N_e N_h}{2d^3} (\br_e - \br_h)^2.
\end{equation}
Here $\br_e$ and $\br_h$ are the oscillator coordinates, and we used 
the fact that in our units $\e^2/\epsilon = 1$. Denoting the base
frequencies of the oscillators by $\omega_e$ and $\omega_h$ we write 
down the coupled equations of motion ($m_e^{*} = 1, m_h^{*} = \kappa$)
\begin{widetext}
\begin{eqnarray}
\label{eq_b2}
  N_e \ddot\br_e + N_e\omega_e^2 \br_e + \frac{N_e e}{c} 
    \dot\br_e\!\times\! \bB
    + \frac{N_e N_h}{d^3} (\br_e - \br_h) = 0,\nonumber\\
  \\ 
  N_h \kappa\ddot\br_h + N_h\kappa\omega_h^2 \br_h - 
    \frac{N_h e}{c} \dot\br_h\!\times\! 
    \bB - \frac{N_e N_h}{d^3} (\br_e - \br_h) = 0.\nonumber
\end{eqnarray}
Eqs.\ (\ref{eq_b2}) are solved by introducing the complex variables
$z_{e(h)} = x_{e(h)} + i y_{e(h)}$ and assuming a harmonic temporal
dependence $z_{e(h)} \sim \exp(i\omega t)$. This leads to the following
secular equation for the resonance frequencies
\begin{equation}
\label{eq_b3}
  \left| \begin{array}{cc}
  -\omega^2 + \omega_e^2 + \omega\omega_c + N_h \Omega_0^2 
   & - N_h\Omega_0^2 \\
  -N_e\Omega_0^2/\kappa & -\omega^2 + \omega_h^2 - \omega\omega_c/\kappa
   + N_e \Omega_0^2/\kappa
  \end{array}\right| = 0,
\end{equation}
\end{widetext}
here we denoted $\omega_c = eB/c$ and $\Omega_0^2 = d^{-3}$. From 
Eq.\ (\ref{eq_b3}) the CM-mode frequencies can be readily obtained as 
solutions of a quartic equation. However, basing on the smallness of
the coupling parameter $\Omega_0^2$, it is possible to extract some 
simpler approximate expressions. We begin by noting that the role of the
terms proportional to $\Omega_0^2$ entering the diagonal and off-diagonal
matrix elements in Eq.\ (\ref{eq_b3}) is diffierent. The former give
first-order corrections to the frequency dispersion, while the latter
contribute to the second order and are important only close to 
coinciding frequencies, thus defining an anticrossing.

Therefore, we first neglect the off-diagonal perturbations and solve
two decoupled quadratic equations originating from the diagonal terms
in Eq.\ (\ref{eq_b3}). The solutions read
\begin{eqnarray}
\label{eq_b4}
  \omega_{1,2} &=& \frac{\omega_c}{2} \pm 
    \sqrt{\frac{\omega_c^2}{4} + \omega_e^2 + N_h\Omega_0^2},\nonumber\\
  \\
  \omega_{3,4} &=& -\frac{\omega_c}{2\kappa} \pm 
    \sqrt{\frac{\omega_c^2}{4\kappa^2} + \omega_h^2 
    + \frac{1}{\kappa} N_e\Omega_0^2}\nonumber.
\end{eqnarray}
The modes of positive frequencies $\omega_1$ and $\omega_3$ obtained 
using the upper signs in Eq.\ (\ref{eq_b4}) are of ``$+$'' circular 
polarisation while the lower-sign solutions $\omega_2$ and 
$\omega_4$ are negative and 
correspond to modes polarised in the ``$-$'' direction. We note, that
Eq.\ (\ref{eq_b4}) predicts that the absolute values of all four
resonance frequencies {\em increase} due to the inter-dot coupling and
gives an estimate of the shifts. In the weak coupling regime they
grow as $\Omega_0^2 \sim d^{-3}$. This conclusion is in agreement with 
our results (see Fig.\ \ref{fig_intro}) and underscores a difference 
of our system from electronic double dots considered in 
Ref.\ \onlinecite{partoens98}. There, all frequency shifts found 
were negative.

Assuming that (as in our numerical calculations) $\omega_e > \omega_h$ 
we find that the modes polarised in ``$-$'' direction will cross. The
crossing point is readily calculated by equating $\omega_2 = \omega_4$ 
and equals
\begin{equation}
\label{eq_b5}
  \omega_{c0} = \frac{\omega_e^2 - \omega_h^2}
  {\sqrt{(1+1/\kappa)(\omega_e^2/\kappa + \omega_h^2)}}.
\end{equation}
Note, that we present the zero-order solution obtained by setting
$\Omega_0^2 \to 0$ which provides an accurate enough estimate. Close 
to this point the off-diagonal perturbations are important and have 
to be taken into account to introduce an anticrossing behaviour.
We use the fact that in the vicinity of the resonances of ``$-$'' 
polarisation (roots $\omega_{2}$ and $\omega_{4}$) the diagonal 
terms of Eq.\ (\ref{eq_b3}) can be approximated by
\begin{eqnarray*}
  -(\omega - \omega_1)(\omega - \omega_2) \approx
    -(\omega_2 - \omega_1)(\omega - \omega_2),\\
  -(\omega - \omega_3)(\omega - \omega_4) \approx
    -(\omega_4 - \omega_3)(\omega - \omega_4),
\end{eqnarray*}
and obtain a quadratic equation 
\[
 (\omega - \omega_2) (\omega - \omega_4) (\omega_1 - \omega_2)
 (\omega_3 - \omega_4) - \frac{N_e N_h}{\kappa} \Omega_0^4 = 0
\]
valid in this region and giving an approximate behaviour of the
anticrossing modes. This equation can be used, in particular, to
estimate the size of the anticrossing gap. To this end we calculate
the difference of its two roots at $\omega_c = \omega_{c0}$ and
$\omega_2 = \omega_4$, and obtain
\begin{eqnarray}
\label{eq_b6}
  \Delta\omega = 2\Omega_0^2 \sqrt{\frac{N_e N_h}
  {\kappa(\omega_1 - \omega_2)(\omega_3 - \omega_4)}},\\
  \nonumber
\end{eqnarray}
with $\omega_{1,2,3,4}$ given by Eq.\ (\ref{eq_b4}).

Equations (\ref{eq_b4}) -- (\ref{eq_b6}) are useful as quick estimates
of essential features (mode shifts, position and size of anticrossing
gap) in the CM-mode spectrum of bipolar quantum dot molecules. These
estimates are obtained for weakly coupled dots and can be asked for 
quantitave predictions only in the limit when inter-dot separation 
considerably exceeds the dot radii. Thus, the gap size obtained from 
Eq.\ (\ref{eq_b6}) agrees with the result of accurate numerical 
calculations within $10~\%$ at $d = 5$. However, as we can see in the 
inset of Fig.\ \ref{fig_cmw} at closer distances the gap grows much more 
slowly than $\sim d^{-3}$ as given by Eq.\ (\ref{eq_b6}).

%


\end{document}